\documentstyle[12pt]{article}
\normalbaselines
\begin{document}

\bibliographystyle{unsrt}
\vbox {\vspace{6mm}}

\begin{center}
{\Large \bf PHOTON DISTRIBUTION FUNCTION FOR STOCKS WAVE FOR 
STIMULATTED RAMAN SCATTERING}
\end{center}

\bigskip

\begin{center}
{\bf O. V.  Man'ko and N. V. Tcherniega} 
\end{center}
\medskip
\begin{center}
{\it P.\,N.\,Lebedev Physical Institute, Russian Academy of Sciences,
Leninsky Pr. 53, Moscow 117924, Russia}
\end{center}

\bigskip
\begin{abstract}
\noindent  New time-dependent integrals of motion are found for 
stimulated Raman scattering. Explicit formula for the photon-number 
probability distribution 
as a function of the laser-field intensity and the medium parameters is 
obtained in terms of Hermite polynomials of two variables.
\end{abstract}

\section{Introduction}

\noindent
 
Since the discovery of stimulated Raman scattering in 1962~\cite{[1]} 
this phenomenon has been intensively investigated both theoretically and 
experimentally~[2--7].
Quantum-mechanical description of stimulated Raman scattering can be 
done in the framework of different equations, namely, Heisenberg--Langeven 
equation~\cite{[7]}, Maxwell--Bloch equation~\cite{[8]}, and Fokker--Planck 
equation~\cite{[9]}. Quantum-statistical properties of stimulated Raman 
scattering were treated using a quadratic Hamiltonian~\cite{[10]}. 
For studying properties of stimulated Raman scattering with taking into 
account intermolecular interaction, a qubic Hamiltonian was used~\cite{[11]}.

Nonclassical properties of stimulated Raman scattering such as squeezing 
and sub-Poissonian statistics were considered in a number of papers~[12--16].
The purpose of our work is to study the photon distribution function for the 
Stocks wave in the framework of the method of linear integrals of motion for 
systems with quadratic Hamiltonians~[17--19]
using the results of~[20--26].
General formulas for matrix elements of the Gaussian density operator for 
the multimode oscillator in the Fock basis were calculated explicitly 
in~\cite{[20]}. For one-mode light described by the Wigner function of a 
generic Gaussian form with five real parameters describing the quadrature 
means, variances, and covariance, the photon distribution function was 
obtained explicitly in terms of Hermite polynomials of two 
variables~\cite{[21]}. In~\cite{[22]}, the temperature dependence of 
oscillations of the photon distribution function for squeezed states was 
investigated. It was shown that oscillations of the photon distribution 
function for squeezed and correlated light were decreasing if the 
temperature increased. 

The photon distribution function for $N$-mode mixed state of light 
described by the Wigner function of the generic Gaussian form was 
calculated explicitly in terms of Hermite polynomials of $2N$ variables 
in~\cite{[23],[24]}, and parameters of the photon distribution function 
were determined through the dispersion matrix and mean values of quadrature 
components of the light. The photon distribution for two-mode squeezed 
vacuum was investigated in~\cite{[25]} where its dependence on four 
parameters (two squeezing parameters, the relative phase between the two 
oscillators, and their spatial orientations) was shown. In~\cite{[26]}, the 
case of generic two-mode squeezed coherent states was considered, and the 
photon distribution function for the states was expressed both through 
four-variable and two-variable Hermite polynomials dependent on two squeezing 
parameters, the relative phase between the two oscillators, their spatial 
orientation, and four-dimensional shift in the phase space of the 
electromagnetic-field oscillator. 

As an application of the theoretical consideration, the linear optical 
transformer of photon statistics for multimode light was suggested~\cite{[27]}.
The transformation coefficient was obtained explicitly in terms of 
multivariable Hermite polynomials. 

In this paper, we find new integrals of motion for the process of stimulated 
Raman scattering. We describe stimulated Raman scattering with the help of 
a simple model of a two-dimensional oscillator, the photons of the Stocks mode 
being described by the one mode of the oscillator and the phonons of the 
medium being described by the another mode of the oscillator. The interaction 
of the photons and phonons is taken to be quadratic in creation and 
annihilation operators of the photons and phonons. We obtain the 
photon--phonon probability distribution function for the Stocks and phonon 
modes after the interaction of the laser field with the 
medium in terms of Hermite polynomials of four variables. The photon 
distribution function for the Stocks mode is found explicitly and expressed 
both in terms of Hermite polynomials of two variables with zero arguments and 
in terms of Legendre polynomials. The mean photon number and its dispersion in 
the Stocks mode are expressed as functions of the medium parameter 
(temperature), the laser frequency, and a parameter of interaction (coupling 
constant).

\section{Integrals of Motion}

\noindent

The simplest phenomenological Hamiltonian, which can be used for the description
of one-mode Stocks-wave excitation, can be written~\cite{[7],[28],[29]}
\begin{equation}\label{eq.1}
\hat H=\hbar\omega_S\hat a^\dagger\hat a +\hbar\omega_{31}\hat b^\dagger 
\hat b +\hbar\kappa\left[e^{-i\omega_L t}\hat a^\dagger\hat b^\dagger
+e^{i\omega_L t}\hat b\hat a\right],
\end{equation}
where $ \hat a$ and $\omega_S $ are the annihilation operator and
the frequency of the Stocks photon, $\hat b$ and $\omega_{31}$ are the 
annihilation operator and the frequency of the phonon, $\omega_L$ is the laser 
frequency, and $\kappa$ is the coupling constant. The laser field is considered
as classical one and its frequency is determined by the condition
\begin{equation}\label{eq.2}
\omega_L=\omega_{31}+\omega_S.
\end{equation}
The damping and depletion of the laser light wave are neglected. 
Antistocks-mode excitation is also neglected and excitation of only one
phonon mode is taken into consideration.

We will show that there exist time-dependent integrals of motion for the model
of Stocks-wave excitation. Let us construct four nonhermitian operators
\begin{eqnarray}
\hat a(t)&=&\hat a e^{i\omega_S t} \cosh \,\kappa t+i\hat b^\dagger
e^{-i\omega_{31}t}\sinh\,\kappa t\,;\nonumber\\
\hat a^\dagger(t)&=&\hat a^\dagger e^{-i\omega_S t} 
\cosh\,\kappa t - i\hat b
e^{i\omega_{31}t}\sinh\,\kappa t\,;\nonumber \\
\label{eq.3}\\
\hat b(t)&=&\hat b e^{i\omega_{31} t} \cosh\,\kappa t
+i\hat a^\dagger e^{-i\omega_S t}\sinh\,\kappa t\,; \nonumber\\
\hat b^\dagger(t)&=&\hat b^\dagger e^{-i\omega_{31} t} 
\cosh\,\kappa t- i\hat a
e^{i\omega_S t}\sinh\,\kappa t\,. \nonumber
\end{eqnarray}
If one introduces two 4-vectors $\mbox {\bf {\cal A}}$ and 
$\mbox {\bf {\cal A}}\left (t\right )$ and uses for the 4-vectors the 
notation
\begin{eqnarray*}   
\mbox {\bf {\cal A}}&=&\left (\hat a,\,\hat b,\,\hat a^\dagger,\,\hat b^\dagger
\right );\nonumber\\ 
\mbox {\bf {\cal A}}\left( t\right )&=&\left (\hat a(t),\,\hat b(t),\,
\hat a^\dagger (t),\,\hat b^\dagger (t)\right ),\nonumber
\end{eqnarray*}                        
the above relations (\ref{eq.3}) may be represented in the matrix form
$$
\mbox {\bf {\cal A}}\left (t\right )=M\left (t\right )\mbox {\bf {\cal A}}.
$$                                    
Here the 4$\times $4-matrix $M\left (t\right )$ is
\begin{equation}\label{M}
M\left (t\right )=\left (\begin{array}{clcr}
e^{i\omega _St}\cosh \,\kappa t&\qquad 0&0&
ie^{-i\omega_{31} t} \sinh \,\kappa t\\
0&e^{i\omega_{31} t} \cosh\,\kappa t
&ie^{-i\omega_S t} \sinh\,\kappa t&0\\
0&-ie^{i\omega_{31} t} \sinh\,\kappa t
&e^{-i\omega_S t} \cosh\,\kappa t&0\\
-ie^{i\omega_S t} \sinh\,\kappa t&\qquad
0&0&e^{-i\omega_{31} t} \cosh\,\kappa t
\end{array}\right ).
\end{equation}
One can see that two components of the vector $\mbox {\bf {\cal A}},$
which are $\hat a,\,\hat b^\dagger $ operators, are transformed independently
of the other two components.                                  

In view of the commutation relations between the photon creation and 
annihilation operators $\hat a,\,\hat a^\dagger$ and the phonon creation and 
annihilation operators $\hat b,\,\hat b^\dagger$, one can check that the 
operators constructed above satisfy boson commutation relations
$$
[\hat a(t),\hat a^\dagger(t)]=1\,;\qquad
[\hat b(t),\hat b^\dagger(t)]=1\,,
$$
and the operators $\hat a(t),~\hat b(t),$ and their hermitian conjugates 
commute as
$$
[\hat a(t), \hat b(t)]=0\,;\qquad [\hat a(t),\hat
b^{\dagger }(t)]=0\,;$$
$$[\hat a^{\dagger }(t),\hat b(t)]=0\,;\qquad 
[\hat a^{\dagger }(t), \hat b^{\dagger }(t)]=0\,.$$
It can be shown, that the total time derivatives 
\begin{eqnarray*}
\frac{d \hat a(t)}{d t}&=&\frac{\partial \hat a(t)}{\partial
t}+\frac{\dot\imath}{\hbar}[\hat H, \hat a(t)]\,;\nonumber\\
\frac{d \hat b(t)}{d t}&=&\frac{\partial \hat b(t)}{\partial
t}+\frac{\dot\imath}{\hbar}[\hat H, \hat b(t)]\nonumber
\end{eqnarray*}
of the operators~(\ref{eq.3}) are equal to zero, i.e.,
\begin{eqnarray*}
\frac{d \hat a(t)}{d t}=0\,;&\qquad &\frac{d \hat a^\dagger(t)}{d
t}=0\,;\nonumber\\
\frac{d \hat b(t)}{d t}=0\,;&\qquad &\frac{d\hat b^\dagger(t)}{d t}=0\,.
\nonumber
\end{eqnarray*}

Consequently, the time-dependent operators 
$\hat a(t),~\hat a^\dagger(t),~\hat b(t),$ and
$\hat b^\dagger(t)$ considered in the Schr\"odinger representation 
are the integrals of motion (linear with respect to the 
photon and phonon creation and annihilation operators) for Stocks-mode 
excitation in the framework of the model with the Hamiltonian~(\ref{eq.1}). 
The operators~(\ref{eq.3}) are equal to standard photon and phonon creation and 
annihilation operators at the initial time moment and their commutators are 
time-independent at all time moments. 

Let us introduce quadrature components of photon and phonon creation and
annihilation operators
\begin{eqnarray*}
\hat p_a=\frac{\hat a -\hat a^\dagger}{i\sqrt2}\,;&\qquad & 
\hat q_a=\frac{\hat a+\hat a^\dagger}{\sqrt2}\,;\nonumber\\
\hat p_b=\frac{\hat b -\hat b^\dagger}{i\sqrt2}\,;&\qquad &
\hat q_b=\frac{\hat b+\hat b^\dagger}{\sqrt2}\,.\nonumber
\end{eqnarray*}
For Stocks-mode excitation, one can write four additional integrals of motion 
using the properties of the integrals of motion~\cite{[18],[19]}
\begin{eqnarray}
\hat p_a(t)&=&\hat p_a\cosh\kappa t\,\cos\omega_S t+\hat q_a\cosh\kappa 
t\,\sin\omega_S t\nonumber\\
&&-\hat p_b\sinh\kappa t\,\sin\omega_{31}t+ \hat q_b\sinh\kappa t\,
\cos\omega_{31}t\,;\nonumber\\
\hat q_a(t)&=&-\hat p_a\cosh\kappa t\,\cos \omega_S t +\hat q_a\cosh\kappa
t\,\cos\omega_S t \nonumber\\
&&- \hat p_b\sinh\kappa t\,\cos\omega_{31}t+\hat q_b
\sinh\kappa t\,\sin\omega_{31}t\,;\nonumber\\
\label{eq.4}\\
\hat p_b(t)&=&-\hat p_a\sinh\kappa t\,\sin\omega_S t+\hat q_a\sinh\kappa t
\,\cos\omega_S t\nonumber\\
 &&+\hat p_b\cosh\kappa t\,\cos\omega_{31}t + \hat q_b\cosh\kappa t\,
\sin\omega_{31}t\,;\nonumber\\
\hat q_b(t)&=&-\hat p_a\sinh\kappa t\,\cos\omega_S t+\hat q_a\sinh\kappa
t\,\sin\omega_S t\nonumber\\
&&-\hat p_b\cosh\kappa t\,\cos\omega_{31}t 
+\hat q_b\cosh\kappa t\,\cos\omega_{13}t\,.\nonumber
\end{eqnarray}
The physical meaning of the invariants~(\ref{eq.4}) is that their
eigenvalues determine the initial values of classical quadrature components in
the phase space of mean values $\langle p_a\rangle ,~\langle p_b\rangle ,
~\langle q_a\rangle ,$ and $\langle q_b\rangle .$ The number of photons does
not conserve in the process of stimulated Raman scattering. But since any
function of integrals of motion is the integral of motion~\cite{[18],[19]}, 
one can find some time-dependent combinations of the photon and phonon 
numbers, which are integrals of motion. Thus, the observable     
\begin{eqnarray*}
N_a(t)&=&\hat a^\dagger (t)\,\hat a(t)\nonumber\\
&=&\hat a^\dagger \hat a\cosh ^2\,\kappa t+
\left(\hat b^\dagger \hat b+1\right)\sinh ^2\,\kappa t\nonumber\\
&&+\frac {i}{2}\left\{\hat a^\dagger \hat
b^\dagger\exp \left[-it\left (\omega _S+\omega _{31}\right )\right ]
-\hat a\hat b\exp \left[it\left (\omega _S+\omega _{31}\right )\right ]
\right \}\sinh \,2\kappa t\nonumber
\end{eqnarray*}                               
is the integral of motion, which has the physical meaning of the initial number
of photons in the system state. The observable
\begin{eqnarray*}
N_b(t)&=&\hat b^\dagger (t)\,\hat b(t)\nonumber\\
&=&\hat b^\dagger \hat b\cosh ^2\,\kappa t+
\left(\hat a^\dagger \hat a+1\right)\sinh ^2\,\kappa t\nonumber\\
&&+\frac {i}{2}\left\{\hat a^\dagger \hat
b^\dagger\exp \left[-it\left (\omega _S+\omega _{31}\right )\right ]
-\hat a\hat b\exp \left[it\left (\omega _S+\omega _{31}\right )\right ]
\right \}\sinh \,2\kappa t\nonumber
\end{eqnarray*}                               
is the integral of motion, which has the physical meaning of the initial number
of phonons in the system state. The difference of the two integrals of motion
$$
N_a(t)-N_b(t)=\hat a^\dagger\hat a -\hat b^\dagger\hat b
$$
is the time-independent integral of motion, which has the physical meaning 
of the difference of photon and phonon numbers in the system, which is 
constant of the motion for the phenomenological Hamiltonian~(\ref{eq.1}). 
Thus, for stimulated Raman scattering we have found new integrals of motion.

\section{Photon--Phonon Probability Distribution Function}

\noindent

In this section, we obtain explicit expression for photon distribution function
of the Stocks mode. Let us introduce the vector column constructed from 
quadrature components of photon and phonon creation and annihilation operators 
in the medium at the initial time moment 
$$
\hat{\mbox{\bf Q}}=(\hat p_a, \hat p_b,\hat q_a, \hat q_b)
$$ 
and the vector column, constructed from the integrals of motion 
$$
\hat{\mbox{\bf I}}\left (t\right )=\left (\hat p_a(t),\hat p_b(t), 
\hat q_a(t),\hat q_b(t)\right ).
$$ 
Then the relation between the integrals of motion~(\ref{eq.4}) and the initial 
quadrature components is of the form 
$$
\hat{\mbox{\bf I}}\left (t\right )=\Lambda(t)\hat{\mbox{\bf Q}}\,,
$$ 
where the real symplectic matrix $\Lambda(t)$ is determined by the equation
\begin{equation}\label{eq.5}
\Lambda(t)=\pmatrix{\cosh \kappa t\cos\omega_S t&-\sinh\kappa t\sin\omega_{13}t
&\cosh\kappa t\sin\omega_S t& \sinh\kappa t\cos\omega_{13}t \cr
-\sinh\kappa t\sin\omega_S t&\cosh\kappa t\cos\omega_{13}t &\sinh\kappa
t\cos\omega_S t& \cosh\kappa t\sin\omega_{13}t \cr -\cosh\kappa t\cos\omega_S t
&-\sinh\kappa t\cos\omega_{13}t&\cosh\kappa t\cos\omega_S t&
\sinh\kappa t\sin\omega_{13}t \cr -\sinh\kappa t\cos\omega_S t &-\cosh\kappa t
\cos\omega_{13}t &\sinh\kappa t\sin \omega_S t& \cosh\kappa t\cos\omega_{13}t
}.  
\end{equation}

Let us introduce the dispersion matrix of quadrature components
$$
\sigma(0)=\pmatrix{\sigma_{p_a^2}& \sigma_{p_a p_b}&\sigma_{p_a q_a}&
\sigma_{p_a q_b} \cr \sigma_{p_a p_b}& \sigma_{p_b^2}& \sigma_{p_b q_a}&
\sigma_{p_b q_b}\cr \sigma_{p_a q_a}& \sigma_{q_a p_b}& \sigma_{q_a^2}&
\sigma_{q_a q_b} \cr \sigma_{p_a q_b}&\sigma_{q_a p_b}& \sigma_{q_a q_b}&
\sigma_{q_b^2} },
$$ 
where the matrix elements are determined through the density matrix as follows 
\begin{eqnarray*}
\sigma_{p_i p_j}&=&\mbox{Tr}\,\hat\rho\hat p_i\hat p_j -
\langle \hat p_i\rangle \langle \hat p_j\rangle ;\nonumber\\
\sigma_{q_i q_j}&=&\mbox{Tr}\,\hat\rho\hat q_i\hat q_j 
-\langle \hat q_i\rangle \langle \hat q_j\rangle ;\nonumber\\
\sigma_{p_i q_j}&=&\frac{1}{2}\mbox{Tr}\,\hat\rho
\left(\hat q_j\hat p_i+\hat p_i\hat q_j \right) -\langle \hat p_i\rangle
\langle \hat q_j\rangle\nonumber
\end{eqnarray*}
(indices $i$ and $j$ can be equal to $a$ and $b$). The dispersion matrix at 
the initial time moment $t$ can be expressed through the initial dispersion 
matrix of quadrature components of medium photons and phonons in the form of 
matrix equation
\begin{equation}\label{eq.6}
\sigma(t)~=\Lambda^{-1}\sigma(0)\Sigma\Lambda^T\Sigma \,,
\end{equation}
where the 4$\times $4-block matrix $\Sigma$ consists of 2$\times $2-zero 
matrices and unity matrices $I_2,$
$$\Sigma~=~\pmatrix{0&I_2 \cr I_2&0}.$$

If the medium photons are in the ground state and phonons are in the state of 
thermodynamical equilibrium with temperature $T$ at the initial time moment, 
then
\begin{eqnarray*}
\sigma_{p_a^2}(0)=\frac {1}{2}\,;&\qquad &\sigma_{q_a^2}(0)=\frac {1}{2}\,;
\nonumber\\
\sigma_{p_b^2}(0)=\frac{1}{2}\coth \frac{\beta}{2}\,;&\qquad &
\sigma_{q_b^2}(0)=\frac{1}{2}\coth \frac{\beta}{2}\,;
\qquad \beta =\frac{\hbar\omega_{31}}{T}\,,\nonumber
\end{eqnarray*} 
and the matrix elements of the matrix $\sigma(t)$ can be found in the 
explicit form
\begin{eqnarray}
\sigma_{p_a^2}(t)&=&\frac{1}{2}\left (\cosh^2\kappa
t+\coth\frac{\beta}{2}\,\sinh^2\kappa t\right );\nonumber\\
\sigma_{p_b^2}(t)&=&\frac{1}{2}\left (\sinh^2\kappa
t+\coth\frac{\beta}{2}\,\cosh^2\kappa t\right );\nonumber\\
\sigma_{p_a p_b}(t)&=&\frac {1}{4}\left(1+\coth \frac {\beta}{2}\right )
\sinh2\kappa t\,\sin\omega_L t\,;\nonumber\\
\sigma_{q_a^2}(t)&=&\cosh^2\kappa t\,\cos\omega_St+\frac {1}{2}\sinh^2\kappa t\,
\coth\frac{\beta}{2}\,;\nonumber\\
\sigma_{q_b^2}(t)&=&\frac {1}{2}\sinh^2\kappa t+ \cosh^2\kappa t\,
\cos^2\omega_{13}t\,\coth\frac{\beta}{2}\,;\nonumber\\
\label{eq.7}\\
\sigma_{q_a q_b}(t)&=&\frac {1}{4}\sinh 2\kappa t\left[\cos\omega_S t\left (
\cos\omega_{13}t-\sin\omega_{13}t\right)
+\coth\frac{\beta}{2}\cos\omega_{13}t
\left (\cos\omega_S t-\sin\omega_S t\right )\right];\nonumber\\
\sigma_{p_a q_a}(t)&=&\frac{\sinh^2\kappa t}{2}\sin2\omega_St\,
\coth\frac{\beta}{2} + \frac{\cosh^2\kappa t}{2}\cos\omega_S t\left (
\cos\omega_S t-\sin\omega_S t\right );\nonumber\\
\sigma_{p_b q_b}(t)&=&\frac{\sinh^2\kappa t}{2}\sin2\omega_{13}t +
\frac{\cosh^2 \kappa t}{2}\coth\frac{\beta}{2}\,\cos\omega_{13}t\left(
\cos\omega_{13}t-\sin\omega_{13}t\right );\nonumber\\
\sigma_{p_a q_b}(t)&=&\frac{1}{4}\sinh2\kappa t\left(
\cos\left(\omega_S-\omega_{13}\right)t +\cos \omega_{13}t\,\coth\frac{\beta}{2}
\left(\sin\omega_S t-\cos\omega_S t\right)\right);\nonumber\\
\sigma_{p_b q_a}(t)&=&\frac{1}{4}\sinh2\kappa t\left(\cos\left(\omega_S
-\omega_{13}\right)t\,\coth\frac{\beta}{2} +\cos\omega_S t\left(
\sin\omega_{13}t -\cos\omega_{13}t\right)\right).\nonumber
\end{eqnarray}
One can see that dispersions of the photon and phonon quadratures become larger
in the process of stimulated Raman scattering. Being initially noncorrelated
the quadratures became statistically-dependent observables, since the 
Hamiltonian~(\ref{eq.1}) is quadratic one.
 
After interacting with the laser field, the state of the system can be 
described by the Wigner function of the Gaussian type
\begin{equation}\label{eq.8}
W(\mbox{\bf Q})=\frac{1}{\sqrt{\mbox{det}\,\sigma(t)}}\,\exp \left(-\frac {1}{2}
\mbox{\bf Q}\sigma^{-1}(t)\mbox{\bf Q}\right), 
\end{equation}
where matrix $\sigma(t)$ is determined by formulas~(\ref{eq.7}). We can express
the inverse matrix $\sigma^{-1}(t)$ at the time moment $t$ through the initial 
inverse matrix $\sigma^{-1}(0)$ using (\ref{eq.6}) and the known property of 
symplectic matrices, namely,
\begin{equation}\label{eq.9}
\sigma^{-1}(t)~=~\Lambda^T\sigma^{-1}(0)\Lambda.
\end{equation}
The matrix elements of the inverse dispersion matrix 
$\sigma^{-1}(t)$~(\ref{eq.9}) have the explicit form
\begin{eqnarray}
\sigma^{-1}_{p_a^2}(t)&=&4\cosh^2\kappa t\,\cos^2\omega_S t +
2\tanh \frac{\beta}{2}\,\sinh^2\kappa t\,;\nonumber\\
\sigma^{-1}_{p_b^2}(t)&=&4\tanh \frac {\beta}{2}\,\cosh^2\kappa t\,
\cos^2\omega_{13}t+2\sinh^2\kappa t\,; \nonumber\\
\sigma^{-1}_{p_a p_b}(t)&=&\sinh2\kappa t\left[\cos\omega_S t
\left(\cos\omega_{13}t-\sin\omega_{13}t\right) 
+\cos\omega_{13}t\,\tanh \frac {\beta}{2}\left (
\cos\omega_S t-\sin\omega_S t\right)\right];\nonumber\\
\sigma^{-1}_{p_a q_a}(t)&=&2\cosh^2\kappa t\,\cos\omega_S t\left(
\sin\omega_St-\cos\omega_S t\right)- 2\tanh\frac {\beta}{2}\,
\sinh^2\kappa t\,\sin2\omega_S t\,;\nonumber\\
\sigma^{-1}_{p_a q_b}(t)&=&\sinh 2\kappa t\left[\cos\omega_S t\,
\cos\omega_{13}t\left(1-\tanh \frac{\beta}{2}\right)
-\sin\omega_{13}\left(\tanh \frac {\beta}{2}\,\sin\omega_S t 
+\cos\omega_S t\right)\right];\nonumber\\
\label{eq.10}\\ 
\sigma^{-1}_{p_b q_a}(t)&=&\sinh2\kappa t\left[\cos\omega_{13}t\,
\cos\omega_S t\left( \tanh \frac {\beta}{2}-1\right)
-\sin\omega_S t\left(\sin\omega_{13}t+\tanh \frac {\beta}{2}\,
\cos\omega_{13}t\right)\right]; \nonumber\\
\sigma^{-1}_{p_b q_b}(t)&=&-2\sinh^2\kappa t\,\sin2\omega_{13}t
+\tanh \frac {\beta}{2}\,\cosh^2\kappa t\left(\sin2\omega_{13}t-
2\cos^2\omega_{13}t\right);\nonumber\\
\sigma^{-1}_{q_a^2}(t)&=&2\left(\cosh^2\kappa t +\tanh \frac {\beta}{2}\,
\sinh^2\kappa t\right);\nonumber\\
\sigma^{-1}_{q_b^2}(t)&=&2\left(\sinh^2\kappa t+\tanh \frac{\beta}{2}\,
\cosh^2\kappa t\right); \nonumber\\ 
\sigma^{-1}_{q_aq_b}(t)&=&\sinh2\kappa t\,\sin \,(\omega_S
+\omega_{13})t\left(1+\tanh \frac {\beta}{2}\right).\nonumber 
\end{eqnarray}

We determine the photon--phonon probability distribution function $P_{n m}$ as 
the probability to obtain $n$ photons in the Stocks mode and $m$ phonons in the
phonon mode after the interaction of the laser field with the medium. The 
function $P_{nm}$ is the matrix element of the density matrix of the system in
the Fock basis  
$$P_{nm }=\langle n, m\mid\hat \rho\mid n, m\rangle ,$$ 
where $\mid n, m\rangle $ is eigenstate of the set of the photon and phonon 
number operators $\hat a^\dagger \hat a$ and $\hat b^\dagger\hat b$
\begin{eqnarray*}
\hat a^\dagger \hat a\mid n,m\rangle &=&n\mid n,m\rangle;\nonumber\\
\hat b^\dagger \hat b\mid n,m\rangle &=&m\mid n,m\rangle.\nonumber
\end{eqnarray*} 
Using the scheme of calculations developed in~[20--26]
for multimode coupled oscillators we arrived at the expression for the 
photon--phonon probability distribution function $P_{nm}$ in terms of Hermite 
polynomials of four variables with zero arguments 
\begin{equation}\label{eq.11}
P_{nm}=\frac{H^{\{\mbox {\bf R}\}}_{nmnm}\left(0,\,0,\,0,\,0\right)}{\left[
\mbox{det}\,\left(\sigma(t)+\mbox {\bf I}_4/2\right)\right]^{1/2}n!\,m!}~, 
\end{equation}
where the matrix $\mbox {\bf R}$ is expressed through the dispersion matrix 
at the time moment~$t$
$$
\mbox {\bf R}=\mbox {\bf U}^\dagger \left(\mbox {\bf I}_4-2\sigma(t)
\right)\left(\mbox {\bf I}_4+2\sigma(t)\right)^{-1}\mbox {\bf U}^*,
$$
the matrix $\mbox {\bf U}$ is
$$
\mbox {\bf U}=\frac{1}{\sqrt2}\pmatrix{-i&0&i&0\cr0&
-i&0&i\cr1&0&1&0\cr0&1&0&1},
$$
and the matrix $\mbox {\bf I}_4$ is four-dimensional unity matrix.

\section{Photon Probability Distribution Function}

\noindent

In experiments, the photon number in the Stocks mode is usually measured. So, 
it is interesting to average the photon--phonon probability distribution 
function over the phonon mode and to obtain the probability to have $n$ 
photons in the Stocks mode.

The photon--phonon probability distribution function can be described by the 
Wigner function (\ref{eq.8}). One has to integrate the Wigner function over the
variables $q_b,~p_b$ in order to obtain the Wigner function (averaged over
the phonon mode) describing the photon state
\begin{eqnarray}\label{eq.12}
W_{\rm {ph}}(q_a,p_a)&=&\frac{1}{2\pi}\int\int_{-\infty}^{\infty} 
W(\mbox{\bf Q})\,dq_b\,dp_b\nonumber\\
&=&\frac{1}{2\pi\sqrt{\mbox{det}\,\sigma(t)}}\int\int_{-\infty}^{\infty}
\exp\left(-\frac{1}{2}\mbox{\bf Q}\sigma^{-1}(t)\mbox{\bf Q}
\right)dq_b\,dp_b\,,
\end{eqnarray}
where $\sigma^{-1}(t)$ is determined by (\ref{eq.9}). The
Wigner function $W_{\rm {ph}}(q_a,p_a)$ (\ref{eq.12}) describes the
photon mode.

It is convinient to change the places of quadrature components. For this 
purpose, we introduce a vector 
$$
\mbox{\bf X}=P\mbox{\bf Q}\,,
$$ 
where the matrix $P$ is of the form
$$
P=\pmatrix{1&0&0&0\cr0&0&1&0\cr0&1&0&0\cr0&0&0&1}.
$$
Then, in view of (\ref{eq.9}), for the argument of exponential function 
in (\ref{eq.12}) one has the equality
$$
\mbox{\bf Q}\sigma^{-1}(t)\mbox{\bf Q}=
\mbox{\bf X}P\Lambda^T\sigma^{-1}(0)\Lambda P\mbox{\bf X}\,.
$$  
By introducing the block matrix 
$$
A=\frac {1}{2}P\Lambda^T\sigma^{-1}(0)\Lambda P=\pmatrix{a&b\cr c&d},
$$
the integral in (\ref{eq.12}) can be rewritten in the form
$$
\int\int_{-\infty}^\infty\exp\left(-\mbox{\bf X} A\mbox{\bf X}\right)
\,d\mbox{\bf y}, \qquad \mbox{\bf y}=(p_b,q_b).
$$  
One can easily see that the integral in (\ref{eq.12}) has the form of the
Gaussian integral calculated in Appendix~A. We obtain, that the Wigner 
function of the photon state of the Stocks mode is described in explicit 
form by the formula
\begin{equation}\label{eq.13}
W_{\rm {ph}}(p_a,q_a)=\frac{1}{2\,\sqrt{\mbox{det}\,\sigma(t)\,\mbox{det}\,
d}}\,\exp{\left[-\frac{1}{2}\pmatrix{p_a& q_a}\sigma^{-1}_{\rm {ph}}
\pmatrix{p_a\cr q_a}\right]},
\end{equation}
where
\begin{equation}\label{eq.14}
\sigma_{\rm {ph}}^{-1}=2a-\frac {1}{2}\left(c^T+b\right)d^{-1}\left(b^T
+c\right)
\end{equation}
with the matrix elements  
$$
\sigma^{-1}_{\rm {ph}}=\pmatrix{s_{11}&s_{12}\cr s_{21}&s_{22}}
$$
expressed through the matrix elements of the inverse photon--phonon matrix 
$\sigma^{-1}(t)$ in the following form    ?????
\begin{eqnarray}\label{eq.15}
s_{11}&=&-\sigma_{p_a^2}^{-1}-\frac{\sigma_{q_b^2}^{-1}\left(\sigma_{p_a
p_b}^{-1}\right)^2 +\sigma_{p_b^2}^{-1}\left(\sigma_{p_a q_b}^{-1}\right)^2
-2\sigma_{p_a p_b}^{-1}\sigma_{p_a q_b}^{-1}\sigma_{p_b
q_b}^{-1}}{2\left[\sigma_{p_b^2}^{-1}\sigma_{q_b^2}^{-1}-\left(\sigma_{p_b
q_b}^2\right)^2\right]}\,; \nonumber\\
s_{22}&=&-\sigma_{q_a^2}^{-1}-\frac{\sigma_{q_b^2}^{-1}\left(\sigma_{p_b
p_a}^{-1}\right)^2 +\sigma_{p_b^2}^{-1}\left(\sigma_{q_a q_b}^{-1}\right)^2
-2\sigma_{q_a q_b}^{-1}\sigma_{p_b q_a}^{-1}\sigma_{p_b
q_b}^{-1}}{2\left[\sigma_{p_b^2}^{-1}\sigma_{q_b^2}^{-1}-\left(\sigma_{p_b
q_b}^2\right)^2\right]}\,; \\
s_{12}=s_{21}
&=&-\sigma_{p_a q_a}^{-1}-\frac{\sigma_{q_b^2}^{-1}\sigma_{p_a
p_b}^{-1}\sigma_{p_b q_a}^{-1} +\sigma_{p_b^2}^{-1}\sigma_{p_a
q_b}^{-1}\sigma_{q_a q_b}^{-1} -\sigma_{p_a q_b}^{-1}\sigma_{p_b
q_a}^{-1}\sigma_{p_b q_b}^{-1}-\sigma_{p_a p_b}^{-1}\sigma_{p_b
q_b}^{-1}\sigma_{q_a q_b}^{-1}}{2\left[\sigma_{p_b^2}^{-1}
\sigma_{q_b^2}^{-1} -\left(\sigma_{p_bq_b}^2\right)^2\right]}\,.\nonumber
\end{eqnarray}

The averaged probability distribution function of photons in the Stocks mode 
can be expressed through the Hermite polynomials of two variables with zero 
arguments using the scheme developed in [20--27]
(see general formulas in Appendix~B)
\begin{equation}\label{eq.16}
P_n =\left[\mbox{det}\left(\sigma_{\rm {ph}}+\frac{\mbox {\bf I}_2}{2}\right)
\right]^{-1/2}\frac{H_{nn}^{\{\tilde {\mbox {\bf R}}\}}(0,0)}{n!}\,,
\end{equation}
where the matrix $\tilde {\mbox {\bf R}}$ is given by the formula
$$ 
\tilde {\mbox {\bf R}}=\mbox {\bf U}^+(\mbox {\bf I}_2-2\sigma_{\rm ph})
(\mbox {\bf I}_2+2\sigma_{\rm {ph}})^{-1}\mbox {\bf U}^* 
$$
with
$$
\mbox {\bf U}=\frac{1}{\sqrt2}\pmatrix{-i&i\cr1&1}.
$$
One can write the photon probability distribution function as a function of
Legendre polynomials using the relations between Hermite and Legendre
polynomials~\cite{[19]}. 

The symmetric matrix $\tilde {\mbox {\bf R}}$ has the form
$$
\tilde {\mbox {\bf R}}=\pmatrix{r_{11}&r_{12}\cr r_{12}&r_{22}}.
$$
Using the relation~\cite{[19]}
$$
H_{nn}^{\{\tilde {\mbox {\bf R}}\}}\left(0,0\right)=
\left (r_{11}r_{22}\right )^{n/2}
H_{nn}^{\{\beta \}}\left(0,0\right),
$$
where the matrix
$$
\beta =\pmatrix{0&r\cr r&0}
$$
has the matrix element
$$
r=\left(r_{11}r_{22}\right)^{-1/2}r_{12},
$$
we arrive at
\begin{equation}\label{eq.17}
P_n=\left[\mbox{det}\,\left(\sigma_{\rm ph}+\frac{\mbox {\bf I}_2}{2}\right)
\right]^{-1/2}(-1)^n\left(r_{12}^2-r_{11}r_{22}\right)^{n/2} 
L_n\left(\frac{r}{\sqrt{r^2-1}}\right), 
\end{equation}
where $L_n$ are Legendre polynomials.

The mean number of photons in the Stocks mode is the function of temperature 
and the coupling constant (of the laser field with the Stocks mode) and it 
can be calculated in explicit form 
\begin{eqnarray}\label{eq.18}
\langle n\rangle&=&\frac{1}{2}\left(\sigma_{p_a p_a}+\sigma_{q_a q_a}
-1\right)\nonumber\\
&=&\frac{1}{2}\left[\cosh^2\kappa t
\left(\frac{1}{2}+\cos^2\omega_S t\right )+ \sinh^2\kappa t
\coth \frac {\beta}{2}-1\right].
\end{eqnarray}
The dispersion of mean photon number in the Stocks mode is
\begin{eqnarray}\label{eq.19}
\sigma_{n^2}(t)&=&\frac {1}{2}\cosh^4\kappa
t\left[\frac {1}{4}+\cos^4\omega_S t+ \cos^2\omega_S t\left(\frac {1}{2}
-\sin\omega_St\right)\right]\nonumber\\ 
&&+\frac {1}{4}\sinh^4\kappa t\coth\frac{\beta}{2}\left(1
+\sin^2 2\omega_S t\right)+\frac {1}{2}\sinh^2\kappa
t\sin2\omega_S t\coth\frac{\beta}{2}\nonumber\\
&&+\frac{1}{8}\sinh^2 2\kappa
t\coth\frac{\beta}{2}\left(\frac {1}{2}+\cos^2\omega_S t\right)\nonumber\\
&&-\frac {1}{2}\cosh^2\kappa t\cos\omega_S t\left(\cos\omega_S t
-\sin\omega_S t\right)-\frac {1}{4}~.
\end{eqnarray}

Thus we have found evolution of the quadrature dispersion matrix due to
photon--phonon interaction. The propagator of the system under consideration in
explicit form is presented in Appendix~C.

\section{Conclusion}

\noindent

We have shown that in the framework of simple quadratic model there exist new
time-dependent integrals of motion for the process of stimulated Raman 
scattering.
The linear (in photon and phonon quadratures) integrals of motion describe the 
initial values of the mean quadratures of the system trajectory in the phase
space. The quadratic (in photon and phonon creation and annihilation operators)
integrals of motion describe the initial numbers of photons and phonons in the 
system state.

Another result of our work is the calculated photon distribution function,
which may be expressed either in terms of Hermite polynomials of two variables
or in terms of Legendre polynomials.

The dependence of the photon number distribution function on the parameters of
the laser field and the coupling constant shows the possibility of processing 
the statistics of stimulated Raman scattering by varying the laser field and 
medium parameters (e.g., temperature). So, the stimulated Raman
scattering can be used for production of nonclassical light. 

Analogous method of investigation can be applied for studying the stimulated 
Brillouin scattering.

\section*{Acknowledgments}

\noindent

The authors are grateful to the International Center for Theoretical Physics
in Trieste for hospitality. The paper was completed during the visit of 
O.V.M. to ICTP as associated member.

The authors thank Professor V.~I.~Man'ko for fruitfull discussions.

The work was partially supported by the Russian Foundation for Basic
Research under Project~No.~96-02-18623.
                                         
\section*{Appendix A}

\noindent
                                     
Here we calculate the Gaussian integral used in Section~4.
We consider the integral in (\ref{eq.12})
$$
\int\exp\left[-{\mbox {\bf X}}A{\mbox {\bf X}}\right]\,dy_1\,\cdots 
\,dy_m\,,
$$
where 
$$
{\mbox {\bf X}}=\left({\mbox {\bf x}},{\mbox {\bf y}}\right)
=\pmatrix{x_1\cr \vdots \cr x_n\cr y_1\cr \vdots \cr y_m}
$$
and the matrix $A$ has four blocks
$$
A=\pmatrix{a&b&\cr c&d}.
$$
We introduce notation
$$
{\mbox {\bf X}}A{\mbox {\bf X}}=\sum _{k=1}^n\sum _{l=1}^nx_ka_{kl}x_l+
\sum_{s=1}^m\sum_{p=1}^my_sd_{sp}y_p
+\sum_{k=1}^n\sum_{s=1}^mx_k\left (b+c^T\right)_{ks}y_s\,.
$$
The matrix $c^T$ is transposed matrix $c.$

Using the general formula for $N$-dimensional Gaussian integral (see, for
example,~\cite{[19]}\,)
$$
\int \cdots \int \exp \left (-{\mbox {\bf Z}}M{\mbox {\bf Z}}
+{\mbox {\bf KZ}}\right )\,dz_1\,\cdots \,dz_N
=\frac{\left(\sqrt\pi\right)^N}{\sqrt {\mbox{det}\,M}}\,\exp\left(
\frac {1}{4}{\mbox {\bf K}}M^{-1}{\mbox {\bf K}}\right),
$$ 
we get 
\begin{equation}\label{eq.20}
\int \exp \left (-{\mbox {\bf X}}A{\mbox {\bf X}}\right )\,d{\mbox {\bf y}}=
\frac{\left(\sqrt\pi\right)^m}{\sqrt{\mbox{det}\,d}}\,\exp\left(-
{\mbox {\bf x}}g{\mbox {\bf x}}\right),
\end{equation}
where the matrix $g$ is
$$
g=a-\frac{1}{4}\left (c^T+b\right )d^{-1}\left (b^T+c\right )
$$
and
$$
d{\mbox {\bf y}}=dy_1\,\cdots \,dy_m.
$$

Using the result obtained one can formulate the following theorem.

Given an arbitrary quantum state of a composed system with a Gaussian Wigner
function $W\left({\mbox {\bf Q}}_1,{\mbox {\bf Q}}_2\right )$ depending on the
set of $2N$ phase-space variables ${\mbox {\bf Q}}_1$ describing the first
subsystem and on the set of $2M$ phase-space variables ${\mbox {\bf Q}}_2$   
describing the second subsystem. Then the Wigner function of the first
subsystem 
$$
W_1\left({\mbox {\bf Q}}_1\right)=\langle W\left({\mbox {\bf Q}}_1,
{\mbox {\bf Q}}_2\right )\rangle _2,
$$
which is the given Wigner function of the composed system averaged over 
variables of the second subsystem ${\mbox {\bf Q}}_2,$  has the
Gaussian form.

The statement takes place for both finite and infinite numbers of the degrees
of freedom $N$ and $M$ of the first and second subsystems.   

\section*{Appendix B}

\noindent

In Appendix~B, we derive general relations of the distribution function
for a system in a Gaussian state to the distribution function of its 
subsystem. The most general mixed squeezed state of the $N$-mode system 
with a {\em Gaussian\/} density 
operator $\hat{\varrho}$ is described by the Wigner function
$W(\mbox{\bf p},\mbox{\bf q})$ of the generic Gaussian form
(see, for example,~\cite{[19],[23]}\,)
\begin{equation}\label{AB1}
W(\mbox{\bf p},\mbox{\bf q})
=(\det \,\sigma )^{-1/2}\exp\left
[-\frac 12\left (\mbox{\bf Q}-\langle\mbox{\bf Q}\rangle \right)
\sigma^{-1}(\mbox{\bf Q}-\langle\mbox{\bf Q}\rangle)\right],
\end{equation}
where $2N$-dimensional vector $\mbox{\bf Q}=(\mbox{\bf p},\mbox{\bf q}
)$ consists of $N$ components $p_1,\ldots ,p_N$ and $N$ components 
$q_1, \ldots ,q_N$, operators $\hat {\mbox{\bf p}}$
and $\hat {\mbox{\bf q}}$ being the 
quadrature components of the creation 
$\hat {\mbox{\bf a}}^\dagger$
and annihilation $\hat {\mbox{\bf a}}$ 
operators (we use dimensionless variables and
assume $\hbar =1$):
$$
\hat {\mbox {\bf p}}=\frac {\hat {\mbox {\bf a}}
-\hat {\mbox {\bf a}}^\dagger}{i\sqrt {
2}}\,;\qquad\hat {\mbox{\bf q}}=
\frac {\hat {\mbox{\bf a}}+\hat {\mbox{\bf a}}^\dagger}{\sqrt {2}}\,.
$$
The $2N$ parameters $\langle p_i\rangle$ and $\langle q_i\rangle $, 
$\,i=1,2,\ldots ,N$, combined into the vector 
$\langle \mbox{\bf Q}\rangle $, are the average values of the quadratures,
$$
\langle \mbox{\bf p}\rangle =\mbox{Tr}~\hat{\varrho}\hat {\mbox{\bf p}}\,;
\qquad \langle\mbox{\bf q}\rangle =\mbox{Tr}~\hat{\varrho}\hat {\mbox{\bf q}},
$$
A real symmetric dispersion matrix $\sigma $ consists of 
($2N^2+N$) variances and covariances
$$
\sigma _{\alpha\beta}=\frac 12\left\langle\hat Q_{\alpha}\hat
Q_{\beta}+\hat Q_{\beta}\hat Q_{\alpha}\right\rangle -\left\langle
\hat Q_{\alpha}\right\rangle\left\langle\hat Q_{\beta}\right\rangle
,\qquad \alpha ,\beta =1,2,\ldots ,2N.
$$
They obey certain constraints, which are nothing but the
generalized uncertainty relations~\cite{[19]}.  From the theorem
formulated in Appendix~A, it follows that the Wigner function of 
a subsystem of the $N$-mode system with $m$ degrees of freedom
$(1\leq m<N)$ has the same form~(\ref{AB1}), namely,
\begin{equation}\label{AB2}
W(\tilde {\mbox{\bf p}},\tilde {\mbox{\bf q}})
=(\det \,\tilde \sigma )^{-1/2}\exp\left
[-\frac 12\left (\tilde{\mbox{\bf Q}}-\langle\tilde{\mbox{\bf Q}}
\rangle \right)\tilde\sigma^{-1}(\tilde{\mbox{\bf Q}}
-\langle\tilde{\mbox{\bf Q}}\rangle)\right],
\end{equation}
where the 2$M$$\times $2$M$-matrix $\tilde \sigma $ is obtained 
from the matrix $\sigma$ by crossing out the rows and columns 
with indices larger than m. The 2$m$-vector $\tilde {\mbox{\bf Q}}$ is
obtained from the vector $\mbox{\bf Q}$ by crossing out components of vectors
$\mbox{\bf p}$ and $\mbox{\bf q}$ with indices larger than $m.$ It means that
\begin{equation}\label{AB3}
W(\tilde{\mbox{\bf p}},\tilde {\mbox{\bf q}})
=(2\pi)^{N-m}\int W(\mbox{\bf p},
\mbox{\bf q})
\,dq_{m+1}\cdots dq_N\,dp_{m+1}\cdots dp_N\,.
\end{equation}
The number distribution function is nothing but the probability  to
have $n_1$ quanta in the first mode, $n_2$ quanta in the second mode,
and so on.  We shall 
designate it by the symbol ${\cal P}_{\mbox{\bf n}}$, where the vector
$\mbox{\bf n}$ consists of $N$ nonnegative 
integers:  $\mbox{\bf n}=(n_1,n_2,\ldots ,n_N)$\mbox{\bf .}  This
probability is given by the formula
$$
{\cal P}_{\mbox{\bf n}}=\mbox{Tr}~\hat{\varrho }\mid\mbox{\bf n}
\rangle \langle\mbox{\bf n}\mid,
$$
where $\hat{\varrho}$ is the density operator of the system under
study, and the state $\mid\mbox{\bf n}\rangle$ is a common eigenstate 
of the set of number operators 
$\hat {a}_i^{\dagger}\,\hat {a}_i,~i=1,2,\ldots ,N$:
$$
\mbox{$\hat {a}_i^{\dagger}\,\hat {a}_i$}\mid \mbox{\bf n}\rangle=
n_i\mid \mbox{\bf n}\rangle.
$$
We use designation
$$
\mbox{\bf n}!=n_1!n_2!\cdots n_N!
$$
and introduce the 2$N$-dimensional unitary matrix
$$
\mbox{\bf U}=\frac 1{\sqrt {2}}\left(\begin{array}{cc}
-i\mbox{\bf I}_N&i\mbox{\bf I}_N\\
\mbox{\bf I}_N&\mbox{\bf I}_N\end{array}
\right).
$$
The symmetric 2$N$-dimensional matrix $\mbox{\bf R}$ and the
$2N$-dimensional vector $\mbox{\bf y},$ given by the relations
\begin{eqnarray}
\mbox{\bf R}&=&\mbox{\bf U}^{\dagger}\left(\mbox{\bf I}_{2N}
-2\sigma\right)\left(\mbox{\bf I}_{2N}+2\sigma\right)^{-1}
\mbox{\bf U}^{*};\\\label{AB6}
\mbox{\bf y}&=&2\mbox{\bf U}^t(\mbox{\bf I}_{2N}-2\sigma )^{-1}\langle
\mbox{\bf Q}\rangle ,\label{AB7}
\end{eqnarray}
determine the number distribution.

The factor ${\cal P}_0$ is nothing but the probability to register no 
quanta.  It equals~\cite{[23]}
$$
{\cal P}_0=\left[\det\left(\sigma+\frac 12\,\mbox{\bf I}_{2N}\right)\right
]^{-1/2}\exp\left[-\langle \mbox {\bf Q}\rangle 
\left(2\sigma+\mbox{\bf I}_{2N}
\right)^{-1}\langle\mbox{\bf Q}\rangle\right].
$$
If 
$$
\langle \mbox{\bf Q}\rangle=\mbox{\bf 0}\,,
$$ 
the probability to have no quanta depends on $(2N-1)$ 
parameters, which coincide up to numerical factors with the
coefficients of the characteristic polynomial of the dispersion
matrix.

The photon distribution function ${\cal P}_{\mbox {\bf n}}$ can be
expressed through the ``diagonal'' multidimensional Hermite
polynomials~\cite{[23]}\,:
\begin{equation}\label{AB9}
{\cal P}_{\mbox{\bf n}}={\cal P}_0\frac {H_{\mbox{\bf n}\mbox{\bf n}}^{
\{\mbox{\bf R}\}}(\mbox{\bf y})}{\mbox{\bf n}!}\,.
\end{equation}
One can derive that
\begin{equation}\label{AB20}
\sigma=\frac 12\left(\mbox{\bf I}_{2N}-\mbox{\bf U}\mbox{\bf R}
\mbox{\bf U}^t\right)\left(\mbox{\bf I}_{2N}
+\mbox{\bf U}\mbox{\bf R}\mbox{\bf U}^t\right)^{
-1}
\end{equation}
and
\begin{equation}\label{B1}
\langle \mbox{\bf Q}\rangle=\frac 12\left(1-2\sigma\right)\left(U^t
\right)^{-1}{\mbox {\bf y}}\,.
\end{equation}

The number distribution function for the subsystem has the same 
form~(\ref{AB9}) but with replacement
\begin{equation}\label{B3}
\tilde {\cal P}_{\mbox{\bf n}}=\tilde {\cal P}_0\frac {H_{\mbox{\bf n}
\mbox{\bf n}}^{\{\tilde {\mbox{\bf R}}\}}(\tilde {\mbox{\bf y}})}
{\mbox{\bf n}!}\,;
\qquad \mbox{\bf n}=\left (n_1,\,n_2,\,\ldots ,\,n_m\right).
\end{equation}
Here
\begin{equation}\label{b4}
\tilde{\cal P}_0=\left[\det\left(\tilde \sigma
+\frac 12\,\mbox{\bf I}_{2m}\right)\right
]^{-1/2}\exp\left[-\langle \tilde{\mbox{\bf Q}}\rangle
\left(2\tilde\sigma+\mbox{\bf I}_{2m}\right)^{-1}\langle\tilde
{\mbox{\bf Q}}\rangle\right].
\end{equation}
The distribution function for a subsystem is completely determined 
by the matrix $\sigma $ and the vector $\mbox{\bf Q}$ since the matrix
$\tilde\sigma $ and the vector $\tilde{\mbox{\bf Q}}$ are determined by 
these quantities.

In view of the physical meaning of the joint distribution functions,
one has the relation
\begin{equation}\label{B5}
\tilde {\cal P}_{\mbox{\bf n}}=\sum_{n_{m+1}=0}^\infty \cdots
\sum_{n_N=0}^\infty {\cal P}_{\mbox {\bf n}}\,.
\end{equation}
Due to this, a sum rule for the multivariable Hermite polynomials
appears, namely,
\begin{equation}\label{B6}
\tilde {\cal P}_0\,{H_{n_1,\ldots ,n_m,n_1,\ldots ,n_m}
^{\{\tilde {\mbox{\bf R}}\}}(\tilde {\mbox{\bf y}})}={\cal P}_{0}
\sum_{n_{m+1}=0}^\infty \cdots \sum_{n_N=0}^\infty 
\frac{H_{n_1,\ldots ,n_N,n_1,\ldots ,n_N}
^{\{\mbox{\bf R}\}}(\mbox{\bf y})}{n_{m+1}!\cdots n_N!}\,.
\end{equation}
Thus, given symmetric 2$N$$\times $2$N$-matrix $R$ and the 2$N$-vector 
$\mbox{\bf y}.$ It means, in view of (\ref{AB20}) and (\ref{B1}), that 
we have the 2$N$$\times $2$N$-matrix $\sigma $ and the 2$N$-vector 
$\langle \mbox{\bf Q}\rangle.$ Using the rule formulated for obtaining 
the 2$m$$\times $2$m$-matrix $\tilde \sigma$ and the 
2$m$-vector $\langle \tilde {\mbox{\bf Q}}\rangle$ from the
matrix $\sigma $ and the vector $\langle \mbox{\bf Q}\rangle,$
we construct the 2$m$$\times $2$m$-matrix $\tilde R$ 
and the 2$m$-vector $\tilde {\mbox{\bf y}}\,:$
\begin{eqnarray}
\tilde{\mbox{\bf R}}&=&\mbox{\bf U}^{\dagger}
\left(\mbox{\bf I}_{2m}
-2\tilde\sigma\right)\left(\mbox{\bf I}_{2m}+2\tilde\sigma\right)^{-1}
\mbox{\bf U}^{*};\\\label{AB6tilde}
\tilde{\mbox{\bf y}}&=&2\mbox{\bf U}^t(\mbox{\bf I}_{2m}
-2\tilde\sigma )^{-1}\langle
\tilde{\mbox{\bf Q}}\rangle .\label{AB7tilde}
\end{eqnarray}
Here $\mbox {\bf U}$ is the 2$m$$\times $2$m$-matrix.

In the left-hand side of the sum rule~(\ref{B6}), the constructed 
arguments and parameters are used. By multiplying the both sides 
of the sum rule~(\ref{B6}) by the factor
$$
\frac {\lambda _1^{n_1}\lambda _2^{n_2}\cdots \lambda _m^{n_m}}
{n_1!\,n_2!\,\cdots n_m!\,}
$$
and doing summation over indices $n_i,$ we arrive at the relation 
for generating functions for quanta distributions of the system and 
the subsystem:
\begin{eqnarray}
G\left(\lambda _1,\ldots ,\lambda _N\right)
&=&\sum_{n_1=0}^{\infty}\cdots\sum_{n_N=0}^{\infty}\frac {\lambda_
1^{n_1}}{n_1!}\frac {\lambda_2^{n_2}}{n_2!}\cdots\frac {\lambda_N^{
n_N}}{n_N!}H_{n_1,n_2,\ldots ,n_N,n_1,n_2,\ldots ,n_N}^{\{\mbox{\bf R}\}}
(\mbox{\bf R}^{-1}\mbox{\bf y})\nonumber\\
&=&\left[\det\left(\Lambda\Sigma_x\mbox{\bf R}
+\mbox{\bf I}_{2N}\right)\right
]^{-1/2}\exp\left[\frac 12\,\mbox{\bf y}\left(\Lambda\Sigma_x\mbox{\bf R}
+\mbox{\bf I}_{2N}\right)^{-1}\Sigma_x\Lambda \mbox{\bf y}\right];\\
\label{B10}
\tilde G\left(\lambda _1,\ldots ,\lambda _m\right)
&=&\sum_{n_1=0}^{\infty}\cdots\sum_{n_m=0}^{\infty}\frac {\lambda_
1^{n_1}}{n_1!}\frac {\lambda_2^{n_2}}{n_2!}\cdots\frac {\lambda_m^{n_m}}{n_m!}
H_{n_1,n_2,\ldots ,n_m,n_1,n_2,\ldots ,n_m}^{\{\tilde{\mbox{\bf R}}\}}
(\tilde{\mbox{\bf R}}^{-1}\tilde{\mbox{\bf y}})\nonumber\\
&=&\left[\det\left(\tilde\Lambda\tilde\Sigma_x\tilde{\mbox{\bf R}}
+\mbox{\bf I}_{2m}\right)\right
]^{-1/2}\exp\left[\frac 12\,\tilde{\mbox{\bf y}}\left(\tilde\Lambda
\tilde\Sigma_x\tilde{\mbox{\bf R}}
+\mbox{\bf I}_{2m}\right)^{-1}\tilde\Sigma_x\tilde\Lambda 
\tilde{\mbox{\bf y}}\right].
\label{B11}
\end{eqnarray}
Here $\mbox{\bf y}=(y_1,\,y_2,\,\ldots ,\,y_{2N})$, the 
2$N$$\times$2$N$-matrix $\Sigma_x$ is the 2$N$-dimensional analog of 
the Pauli matrix $\sigma_x$,
$$
\Sigma_x=\left(\begin{array}{cc}
0&\mbox{\bf I}_N\\
\mbox{\bf I}_N&0\end{array}
\right)\equiv \mbox{\bf U}^{\dagger}\mbox{\bf U}^{*}=\mbox{\bf U}^t
\mbox{\bf U}\,,
$$
and the diagonal 2$N$$\times$2$N$-matrix $\Lambda$ reads
$$
\Lambda =\sum_{j=1}^N\lambda_j\Lambda_j,
$$
where each matrix $\Lambda_j$ has only two nonzero elements:
$$
(\Lambda_j)_{jj}=(\Lambda_j)_{j+N,\,j+N}=1.
$$
The matrices $\tilde\Sigma_x$ and $\tilde \Lambda _x$are defined by
analogous formulas but in 2$m$-dimensional space.

The relation for the generating functions is
\begin{equation}\label{B12}
\tilde {\cal P}_0\,\tilde G\left(\lambda _1,\ldots ,\lambda _m\right)
={\cal P}_0\,G\left(\lambda _1,\ldots ,\lambda _m,1,1,\ldots,1\right).
\end{equation}

We have shown that for any Gaussian state (a state with a Gaussian 
Wigner function) of $N$-mode system, the number distribution function
of a $m$-mode subsystem is described by the diagonal Hermite polynomials
with 2$m$ indices. The parameters of these polynomials are determined 
explicitly by quadrature dispersion matrix and quadrature means of the
$N$-mode system. The discussed model of the stimulated Raman scattering
is the two-mode example of the suggested construction for its partial 
case corresponding to zero values of the quadrature means.

\section*{Appendix C}

\noindent
          
In this Appendix, we present the propagator of the two-mode system containing
photons and phonons in coordinate representation. We introduce four 
2$\times $2-matrices 
\begin{eqnarray*}
\lambda _1&=&\pmatrix{\cosh \kappa t\cos\omega_S t&
-\sinh\kappa t\sin\omega_{13}t\cr
-\sinh\kappa t\sin\omega_S t &\cosh\kappa t\cos\omega_{13}t};\nonumber\\ 
\lambda _2&=&\pmatrix{\cosh\kappa t\sin\omega_S t& 
\sinh\kappa t\cos\omega_{13}t\cr
\sinh\kappa t\cos\omega_S t&
\cosh\kappa t\sin\omega_{13}t};\nonumber\\ 
\lambda _3&=&\pmatrix{-\cosh \kappa t\cos\omega_S t&
-\sinh\kappa t\cos\omega_{13}t\cr
-\sinh\kappa t\cos\omega_S t &-\cosh\kappa t\cos\omega_{13}t};\nonumber\\
\lambda _4&=&\pmatrix{\cosh \kappa t\cos\omega_S t&
\sinh\kappa t\sin\omega_{13}t\cr
\sinh\kappa t\sin\omega_S t &\cosh\kappa t\cos\omega_{13}t}.\nonumber
\end{eqnarray*}
The matrix~(\ref{eq.5}) determining the linear integrals of motion 
$\hat {\mbox {\bf I}}\left(t\right)$ is expressed in terms of these 
2$\times $2-matrices 
$$
\Lambda \left (t\right )=\left (\begin{array}{clcr}
\lambda _1&\lambda _2\\
\lambda _3&\lambda _4\end{array}\right ).                                   
$$                                       
Then, in view of general formula for the propagator of quadratic 
systems~\cite{[19]}, $G\left (\mbox {\bf x}_1,\,\mbox {\bf x}_2,\,t\right )$ 
reads in coordinate representation  
$$
G\left (\mbox {\bf x}_1,\,\mbox {\bf x}_2,\,t\right )=\left [\mbox {det}\,
(-2\pi i\lambda _3)\right ]^{-1/2}\exp \left \{-\frac {i}{2}\left [
\mbox {\bf x}_2\lambda _3^{-1}\lambda _4\mbox {\bf x}_2
-2\mbox {\bf x}_2\lambda _3^{-1}\mbox {\bf x}_1
+\mbox {\bf x}_1\lambda _1\lambda _3^{-1}\mbox {\bf x}_1\right ]\right \},
$$
where $\mbox {\bf x}_1=\left (x_{11},\,x_{12}\right )$ and
$\mbox {\bf x}_2=\left (x_{21},\,x_{22}\right )$ are the initial and final
quadrature components of the photon and phonon modes.

Thus the Green function of the system under consideration is constructed.


\begin{thebibliography}{99}

\bibitem{[1]} 
G.~Eckhard, R.~W.~Hellwarth, F.~J.~McClung, et al., {\sl Phys. Rev. Lett.} A, 
{\bf 9}, 455 (1962).
 
\bibitem{[2]} 
N.~Bloembergen, {\sl Am. J. Phys.}, {\bf 35}, 989 (1967). 

\bibitem{[3]} 
A.~Z.~Grasyuk, {\it Lasers and their Applications}, {\sl Proceedings of the 
Lebedev Physical Institute}, Nauka, Moscow (1986), Vol.~86.

\bibitem{[4]} 
C.~S.~Wong, {\sl Phys. Rev.}, {\bf 182}, 482 (1969).

\bibitem{[5]} 
S.~Kielich, {\sl Prog. Opt.}, {\bf 20}, 156 (1983).

\bibitem{[6]} 
Y.~R.~Shen, {\it The Principles of Nonlinear Optics}, Wiley, 
New York (1984).

\bibitem{[7]} 
M.~G.~Raymer and I.~A.~Walmsley, {\sl Prog. Opt.}, {\bf 28}, 182
(1990). 

\bibitem{[8]} 
A.~Schenzle and H.~Brandt, {\sl Phys. Rev.} A, {\bf 20}, 1928 (1979). 

\bibitem{[9]} 
R.~Loudon, {\it The Quantum Theory of Light}, Clarendon Press, Oxford 
(1983). 

\bibitem{[10]} 
E.~A.~Mishkin and D.~F.~Walls, {\sl Phys. Rev.}, {\bf 185}, 1618 (1969).

\bibitem{[11]} 
A.~Miranovicz and S.~Kielich, {\it Modern Nonlinear Optics}, Willey,
New York (1994), Vol.~LXXXV.

\bibitem{[12]} 
J.~Perina, V.~Perinova, and  J.~Kodousek, {\sl Opt. Comm.}, {\bf 49}, 210
(1994). 
 
\bibitem{[13]} 
V.~Perinova, M.~Karska, and J.~Krepelka, {\sl Acta Phys. Pol.} A, {\bf 79},
817 (1991).  

\bibitem{[14]} 
B.~K.~Mohanti, N.~Nayak, and P.~S.~Gupta, {\sl Opt.  Acta}, {\bf 29},
1017 (1982).

\bibitem{[15]} 
P.~S.~Gupta and J.~Dush, {\sl Czech. J. Phys.}, {\bf 40}, 432 (1990).

\bibitem{[16]} 
H.~Ritsch, M.~A.~M.~Marte, and P.~Zoller, {\sl Europhys. Lett.}, {\bf 19},
7 (1992). 

\bibitem{[17]} 
I.~A.~Malkin, V.~I.~Man'ko, and D.~A.~Trifonov, {\sl J.  Math. Phys.}, 
{\bf 14}, 576 (1973).

\bibitem{[18]} 
I.~A.~Malkin and V.~I.~Man'ko, {\it Dynamical Symmetries and Coherent States 
of Quantum Systems} [in Russian], Nauka, Moscow (1979).

\bibitem{[19]} 
V.~V.~Dodonov and V.~I.~Man'ko, {\it Invariants and Evolution of Nonstationary 
Quantum Systems}, {\sl Proceedings of the Lebedev Physical Institute}, Nova 
Science, New York (1989), Vol.~183.

\bibitem{[20]} 
V.~V.~Dodonov, V.~I.~Man'ko, and V.~V.~Semjonov, {\sl Nuovo Cim.} B, {\bf 83},
145 (1984).

\bibitem{[21]} 
V.~V.~Dodonov, O.~V.~Man'ko, and V.~I.~Man'ko, {\sl Phys. Rev.} A, {\bf 49},
2993 (1994). 

\bibitem{[22]} 
V.~V.~Dodonov, O.~V.~Man'ko, V.~I.~Man'ko, and L.~Rosa, {\sl Phys.  Lett.} A, 
{\bf 185}, 231 (1994).

\bibitem{[23]} 
V.~V.~Dodonov, O.~V.~Man'ko, and V.~I.~Man'ko, {\sl Phys. Rev.}
A, {\bf 50}, 813 (1994).

\bibitem{[24]} 
V.~V.~Dodonov, O.~V.~Man'ko, and V.~I.~Man'ko, {\sl Bulletin of the Lebedev 
Physical Institute} (Allerton, New York), No.~4, 5 (1994).

\bibitem{[25]} 
G.~Schrade, V.~Akulin, V.~I.~Man'ko, and W.~Schleich, {\sl Phys.
Rev.} A, {\bf 48}, 3854 (1991).

\bibitem{[26]} 
G.~Schrade, O.~V.~Man'ko, V.~I.~Man'ko, and W.~Schleich, ``Photon Statistics 
of Generic Two-Mode Squeezed Coherent Light'' (in preparation).  

\bibitem{[27]} 
V.~V.~Dodonov, O.~V.~Man'ko, V.~I.~Man'ko, and P.~G.~Polynkin, {\sl J. Russ.
Laser Research}, {\bf 17}, 449 (1996). 

\bibitem{[28]} 
D.~F.~Walls, {\sl Z. Physik}, {\bf 237}, 224 (1970).

\bibitem{[29]} 
D.~F.~Walls, {\sl Z. Physik}, {\bf 234}, 331 (1970). 


\end{thebibliography}
\end{document}